# Evolutionary Predictability and Complications with Additivity

Kristina Crona, Devin Greene, Miriam Barlow

University of California, Merced

5200 North Lake Road

Merced, CA 95343

**Abstract**

Adaptation is a central topic in theoretical biology, of practical importance for analyzing drug resistance mutations. Several authors have used arguments based on extreme value theory in their work on adaptation. There are complications with these approaches if fitness is additive (meaning that fitness effects of mutations sum), or whenever there is more additivity than what one would expect in an uncorrelated fitness landscapes. However, the approaches have been used in published work, even in situations with substantial amounts of additivity. In particular, extreme value theory has been used in discussions on evolutionary predictability. We say that evolution is predictable if the use of a particular drug at different locations tends lead to the same resistance mutations. Evolutionary predictability depends on the probabilities of mutational trajectories. Arguments about probabilities based on extreme value theory can be misleading. Additivity may cause errors in estimates of the probabilities of some mutational trajectories by a factor 20 even for rather small examples. We show that additivity gives systematic errors so as to exaggerate the differences between the most and the least likely trajectory. As a result of this bias, evolution may appear more predictable than it is. From a broader perspective, our results suggest that approaches which depend on the Orr-Gillespie theory are likely to give misleading results for realistic fitness landscapes whenever one considers adaptation in several steps.

**1. Introduction**

1.1 Evolutionary predictability

Given a recent change in the environment, such as the use of antibiotics, a population may change through a sequence of mutations before it reaches a new fitness optimum, such as a mutant variant with a high level of antibiotic resistance. The original genotype is referred to as the wild-type. Adaption processes are of theoretical and practical interest. A classical metaphor for an intuitive understanding of adaptation is the genotypic fitness landscape (Wright, 1931). Informally, adaptation can be pictured as an uphill walk in the fitness landscapes, where height represent fitness and where each step is between genotypes which differ by exactly one point mutation. In terms of the metaphor, consider a landscape with a single peak, so that each uphill walk from the wild-type eventually leads to the peak. Sometimes there are several possible walks to the peak. Evolution may still be fairly predictable if most walks are very unlikely indeed, so that nature will almost certainly chose among the remaining few walks. A slightly different question concerns landscapes with several peaks, where one asks if nature is more likely to chose some peaks than others. In general, evolutionary predictability depends on the probabilities of mutational trajectories.

Throughout the paper, we will focus on evolutionary predictability for single peaked landscapes. In addition to the theoretical interest, predictability is sometimes of clinical importance. For instance, in order to manage drug resistance it may be useful to know the order in which mutations are likely to accumulate. Sometimes a mutation may be selected for only if a different mutation has already occurred (Desper et al., 1999; Beerenwinkel et al., 2007).

The influential Orr-Gillespie approach to adaptation (e.g. Orr, 2002) has been used in discussions about evolutionary predictability (Weinreich et al., 2006). In particular, the authors argue that evolution is much more predictable than has so far been appreciated, and employ extreme value theory to provide intuition for the adaptation process. We will discuss the approach in Weinreich et al. (2006), and similar approaches which depend on extreme value. From a broader perspective, we are interested in how methods which are suitable for uncorrelated fitness landscapes would perform for realistic fitness landscapes. By definition, for an uncorrelated fitness landscapes, there is no correlation between the fitness of a genotype and the fitness of its mutational neighbors, i.e., genotypes that differ by one mutation only. Uncorrelated fitness landscapes have been studied extensively (e.g. Kingman, 1978; Kauffman and Levin, 1987; Flyvberg and Lautrup, 1992; Macken and Perelson, 1989; Orr 2002, Rokyta et al., 2006; Park and Krug, 2008). Consequently, it is clearly of interest whether results on uncorrelated fitness could apply to realistic fitness landscapes, at least as approximations. We focus on evolutionary predictability, where we will carefully test predictions based on the Orr-Gillespie approach.

The paper is structured as follows. We first describe one of the estimates we will study, and discuss why this type of estimates could be useful in principle. Then we give some background about fitness landscapes. The result section describes estimates which depend on the Gillespie-Orr approach, followed by tests of how the estimates perform, and a proposal of an alternative approach which does not depend on the Orr-Gillespie theory. Finally, our conclusions are summarized in a discussion section.

*1.2 An estimate found in empirical work*

Formula S5b in Weinreich et al. 2006 concerns fixation probabilities of beneficial mutations, and was used is a discussion about cefotaxime resistance. We will test how well the formula predicts probabilities of mutational trajectories for realistic fitness landscapes. More precisely, we will consider the following estimate of probabilities, based on Formula S5b. Notice that the formula depends on fitness ranks of the alleles only and, by convention, the genotype with the highest fitness has rank 1.

**Estimate based on Formula S5b:**

$$\frac{\sum_{x=r_j}^{r_i-1} \frac{1}{x}}{\sum_{k \in M_i^+} \sum_{x=r_k}^{r_i-1} \frac{1}{x}}$$

where $M_i^+$ denotes the set of mutational neighbors of genotype i that are beneficial. The formula in the estimate above is identical with S5b in Weinreich et al. (2006), and the support for it consists of simulations. However, we are not convinced that the simulations make sense for realistic fitness landscape. In order to investigate this problem, we will test how well Formula S5b predicts probabilities of mutational trajectories.

Estimates of [reproductive] fitness of genotypes are obviously interesting for theoretical reasons because the connection to evolutionary predictability. However, fitness estimates based on fitness ranks of genotypes could be of practical interest as well. Indeed, it may be complicated and costly to measure fitness. Moreover, sometimes it is difficult or even impossible to measure fitness with reasonable accuracy in nature. Information about fitness ranks of genotypes may nevertheless be available.

Consider for instance drug resistance. Disk diffusion tests provide information about fitness ranks, whereas more fine scaled fitness differences are not revealed (e.g. Jorgensen, J. H. and Ferraro, M. J. , 2009). The most precise methods for quantifying fitness differences are considerably more expensive than disk diffusion tests. Moreover, even if one can measure fitness, laboratory measurements do not always agree with clinical findings. Consequently, one may prefer clinical information of qualitative type to laboratory data. The theme of utilizing indirect information about fitness ranks is developed in Crona et al. (2013b), see also Crona et al. (2013a); Crona (2013). In order to take full advantage of available empirical data, for instance mutation records, it would obviously be useful with fitness estimates based on fitness ranks. Before discussing such estimates in more detail, we will give some background about fitness landscapes.

*1.3 Fitness landscapes, uncorrelated fitness and extreme value theory*

An additive fitness landscape, or a non-epistatic landscape, has the property that the fitness effect of beneficial mutations sum. Suppose that the wild-type ab has fitness 1, the genotype Ab has fitness 1.01, and the genotype aB has fitness 1.02. Then the genotype AB has fitness 1.03. (In the literature non-epistatic fitness is sometimes defined as multiplicative, so that AB would have fitness 1.0302. If the fitness effects and the number of beneficial mutations are small, there is not much difference between the definitions.) In contrast, as mentioned there is no correlation between the fitness of a genotype and the fitness of its mutational neighbors for an uncorrelated (rugged or random) fitness landscape. In particular, consider a genotype of some mean fitness. If two single mutants correspond to higher fitness, then the corresponding double mutant is equally likely to be less fit as it is to be more fit than the original genotype. Uncorrelated fitness and additivity can be considered as two extremes with regard to the amount of structure in the fitness landscape, and many fitness landscape fall between the extremes.

Recent findings indicate that empirical fitness landscapes tend to have considerably more additivity than what one would expect from an uncorrelated fitness landscape, including in cases with epistasis and sign epistasis (e.g. Szendro et al., 2013; Goulart et al. 2013; Kouyos et al. 2012; Carnerio and Hartl, 2010). In other words, if a landscape deviates from being additive, it may still have considerably more additivity than expected for an uncorrelated fitness landscape.

The Orr-Gillespie approach to adaptation depends on extreme value theory for estimating fitnesses of beneficial mutations (e.g. Orr, 2002), and has been used in discussions about evolutionary predictability. We will investigate errors in fitness estimates based on fitness ranks, with focus on how the errors may influence conclusions about evolutionary predictability

Throughout the paper, we assume that the population is monomorphic most of the time, and that a beneficial mutation goes to fixation before the next mutation occurs. More precisely, we assume the strong selection weak mutation regime (SSWM) (Gillespie 1983, 1984; Maynard Smith 1970).

We will make references to a frequently used result (e.g. Orr, 2002) regarding fitness spacing for beneficial mutations. In brief, assume that fitnesses for genotypes are drawn from some probability distribution of the Type III class, a class that includes exponential, gamma, normal and log-normal distributions. Under the assumption that the fitnesses of the wild-type and the more fit genotypes belong to the extreme right tail of the fitness distribution, one can apply limiting results from extreme value theory that describe the tail behavior of such distributions in order to predict the fitness spacing of these genotypes.
In particular, if one assumes that the wild-type has k mutational neighbors, which are more fit than the wild-type, then after ordering the k mutational neighbors by decreasing fitness, the difference between the fitness of the first and the second is expected to be greater than the difference between the second and the third, etc. More precisely, the fitness differences between the first and the second genotype, the second and the third genotype, etc. are expected to be distributed as

$$c, \frac{c}{2}, \frac{c}{3}, \ldots, \frac{c}{k} \qquad [1],$$

for some constant c.  Elementary statistical considerations show that all estimates based on [1] will give very rough approximations even under ideal circumstances, but one can say that [1] constitutes our best guess.

For more background regarding the assumptions described, see e.g. Orr (2002); Gillespie (1983, 1984). For more background about fitness landscapes relevant in this context, see e.g. Kryazhimskiy et al. (2009); Szendro et al.(2013b); Crona (2013), Crona et al. (2013). For a review article on evolutionary predictability, see Lobkovsky and Koonin (2012), and for recent work on the same topic, see e.g. Szendro et al. (2013a); Ostman and Adami (2013); Dobler (2012);Lobkovsky et al. (2011).

## 2. Results

*2.1 Probabilities of mutational trajectories*

By the SSWM assumption we consider fixation of beneficial mutations as independent events, so the probability for a trajectory equals the product of the probabilities of its steps. To find the probabilities for single substitutions we used the following well-established estimate:

The probability that a beneficial mutation j will be substituted at the next step in adaptation is:

$$\frac{s_j}{s_1 + \ldots + s_k} \quad [2],$$

where $s_r$ the fitness contribution of mutation *r*, and where there are *k* beneficial mutations in total. This estimate is appropriate when the fitness contribution of each mutation is small.

Suppose that there are exactly two beneficial mutations for some wild-type. Moreover, assume that the fitness spacing agrees with the expected average fitnesses predicted by extreme value theory (see the list [1]). Then the fitness contributions are

$$w_0 + c + \frac{c}{2}, \quad w_0 + \frac{c}{2},$$

where $w_0$ is the fitness of the wild-type and where c is a constant. Indeed, this condition means exactly that the corresponding fitness differences

$$(w_0 + c + \frac{c}{2}) - (w_0 + \frac{c}{2}) = c, \quad (w_0 + \frac{c}{2}) - w_0 = \frac{c}{2}.$$

agree with [1]. We call such a set of mutation an **E₂** set. We define an **E₃** set similarly so that [1] is satisfied for three beneficial mutations.

**Example 1** Consider an **E₃** set where fitness is additive. Notice that the triple mutant is the fittest variant that can be obtained from a subset of the three mutations. From these assumptions it is straight forward to compute probabilities for the mutational trajectories.

By assumption, the three mutations contributed to fitness according to the following list:

$$1: c + \frac{c}{2} + \frac{c}{3} = \frac{11c}{6}$$
$$2: \frac{c}{2} + \frac{c}{3} = \frac{5c}{6}$$
$$3: \frac{c}{3}$$

Using standard notation, the fitnesses of the genotypes are as follows.:

$$w(g_{000})=w_0, \quad w(g_{100})=w_0+\frac{11c}{6}, \quad w(g_{010})=w_0+\frac{5c}{6}, \quad w(g_{001})=w_0+\frac{c}{3},$$
$$w(g_{110})=w_0+\frac{16c}{6}, \quad w(g_{101})=w_0+\frac{13c}{6}, \quad w(g_{011})=w_0+\frac{7c}{6}, \quad w(g_{111})=w_0+3c.$$

There are 3!=6 mutational trajectories from the wild-type g000 to the triple mutant g111 . As explained, the probability of a particular trajectory is the product of the probabilities of each step of the trajectory by assumption, so that [2] implies that the probability of the trajectory A : g000 →g100 → g110 →g111 is

$$\frac{11}{11+5+2} \cdot \frac{5}{5+2} \cdot 1 = 0.437$$

Similarly, the probability of the trajectory B : g000 → g001 → g011 → g111 is

$$\frac{3}{11+5+2} \cdot \frac{5}{11+5} \cdot 1 = 0.0521$$

The average probability of a trajectory from the wild-type to the triple mutant is 1/6 = 0.17. Trajectory A has the highest probability (0.437) and trajectory B has the lowest probability (0.0521). Trajectory A is eight times more likely than trajectory B. This shows that the probabilities for trajectories are unevenly distributed.

2.2 *Estimates based on extreme value theory.*

Consider a *n* beneficial mutations. Assume that the fitness spacing agrees with the expected average fitnesses predicted by extreme value theory (see the list [1]). We will call such a set of n mutation an $E_n$ set. More precisely an $E_n$ set consists of *n* beneficial mutations with fitness contributions (where "1:", "2: " and so forth are followed by he fitness of the genotype of fitness rank 1, 2 and so forth);

$$1:\sum_{i=1}^{n}\frac{c}{i},\quad 2:\sum_{i=2}^{n}\frac{c}{i},\quad 3:\sum_{i=3}^{n}\frac{c}{i},\ldots,\quad n-1:c\left(\frac{1}{n-1}+\frac{1}{n}\right),\quad n:\frac{c}{n},$$

where *c* is a small constant. The condition described is equivalent to that the fitness differences agree with Equation [1]. For our tests of fitness estimates, we will study $E_n$ sets assuming that fitness is additive, which will be our focus throughout the paper. Our motivation for testing fitness estimates on this particular case, will be given later. As demonstrated in Example 1, we can compute the probabilities for the n! mutational trajectories from the wild-type to the n-tuple mutant which constitutes the new fitness optimum.

The following procedures for estimating probabilities of mutational trajectories, or closely related arguments, have been used in published work. For both estimates, the input is fitness ranks of genotypes only. Recall that the genotype with highest fitness has rank 1.

**Estimate 1.** The first estimate relates to approaches in Orr, (2002). If k beneficial mutations are available, we estimate the probability for a particular mutation, by assuming that the *k* mutations constitute an $E_k$ set, and that Equation [2] holds.

Notice that the probabilities of particular mutations are determined by the two assumptions stated in the estimate.

From now on, we will refer to the estimate based on S5b, Weinreich et al. (2006), as described in the introduction, as "Estimate 2".

**Estimate 2.** Consider the wild-type, an n-tuple mutant and all intermediates. Rank the $2^n$ intermediate genotypes according to fitness. Let $r_k$ represent the fitness rank of genotype *k* among all $2^n$ intermediate genotypes, regardless of mutational adjacency.
The probability that the next step in the trajectory is genotype *j*, when the present genotype is *i*, is given by the formula

$$\frac{\sum_{x=r_j}^{r_i-1} \frac{1}{x}}{\sum_{k \in M_i^+} \sum_{x=r_k}^{r_i-1} \frac{1}{x}}$$

where $M_i^+$ denotes the set of mutational neighbors of genotype i that are beneficial.

The reader may wonder what the justification for using Estimate 1 and 2, or related approaches based on extreme value theory, could be. For clarity, the authors of this paper have never used or recommended Estimate 1 or 2. By inspection, it seems to that this type of estimates could be reasonable for uncorrelated fitness landscapes, assuming the results from extreme value theory discussed in the introduction. However, the estimates are always coarse. More important, it is not clear how well the estimates perform for empirical fitness landscapes because of the complication with additivity. It is important to note that the formula in Estimate 2 was used in an empirical study (S5b, Weinreich et al., 2006). As far as we know, Estimate 1 has not been used in empirical work.

Our reasons for including Estimate 1 in this study are two-fold. For readers interested in analyzing the underlying factors for errors in Estimate 2, it is useful to have Estimate 1 as a comparison. Estimate 1 is in a sense more intuitive. Moreover, since we are interested in understanding how arguments based on the Orr-Gillespie approach handle additivity in general, it is useful to have more than one single example.

*2.3 A test case for the estimates*

In order to quantify the error produced by Estimates 1 and 2, we consider an $\mathbf{E_n}$ set of mutations. Moreover, we assume that the fitness effects of the mutations in $\mathbf{E_n}$ sum. It follows that there are exactly n! trajectories from the wild-type to the n-tuple mutant at the fitness optimum. This assumption along with the SSWM assumption defines the situation we consider. Our assumptions specify a good test case for Estimates 1 and 2. Informally, the landscapes are "unusually typical". Indeed, the fitness spacing agrees exactly with the expected average behavior predicted by extreme value theory. Moreover, we consider additive fitness a good a test case. Indeed, if the estimates are accurate for empirical fitness landscapes, it is reasonable to expect that the estimates perform well for additive fitness landscapes.

A comment is probably necessary for readers who do not immediately understand the meaning of a "test case" or a "counterexample". A good scientific model or method should work under very general assumptions. Consequently, a test case could be rather special and if the general method fails or the test case, one has a counterexample. Counterexamples are rarely used in modern biology, but the power of a single counterexample is enormous (if an apple went straight up when dropped, one would have to question Newton's laws of universal gravitation.)

A common misconception is that a counterexample is intended as an alternative theory or model. However, a counterexample is merely a case where a prediction from a general theory went wrong. Consequently, It is perfectly fine that a counterexampel is special. After this explanation, the reader should be prepared for considering our test case, which admittedly is somewhat special.

Consider n= 3 (Example 1), but under the assumption that we only have knowledge about the fitness ranks of the genotypes in Example 1. From that information we cannot determine the probabilities of the trajectories. Suppose that we use Estimate 1 for an approximation of the probability of trajectory A. That would estimate the probability that each transition is to the genotype of greatest fitness among available alternatives. According to Estimate 1, we assume an $E_2$ set, so that the probability for the first step is

$$\frac{11}{11+5+2}$$

For the next step, according to Estimate 1, we assume an $E_2$ set so that the probability for the next transition is 3/4.

It follows that Estimate 1 of the probability for trajectory A is

$$\frac{11}{11+5+2} \cdot \frac{3}{4} \cdot 1 = 0.458$$

(or 1.05 times the correct value 0.437). It is clear why Estimate 1 gives an overestimate. The fitness contributions 3c/2 and c/2 of the genotypes s in **E₂** differ by a factor 3, whereas the fitness contributions of the two least fit genotypes s in **E₂** differ by a factor 2.5.

In order to investigate the accuracy of Estimate 1 and Estimate 2 we consider the probabilities for several mutational trajectories for n = 4, 5, 6. We enumerate the mutations by decreasing fitness, and label the trajectories according to the order in which the mutations occur.

*2.4 The two extreme trajectories for the test case*

For any n ≥ 3 consider the trajectory where each transition is to the genotype of greatest fitness among available alternatives (such as trajectory A in Example and the trajectory where each transition is to the genotype of smallest fitness among available alternatives (such as trajectory B in Example 1). The first trajectory is denoted Trajectory 12 · · · n, and the second trajectory is denoted Trajectory n(n - 1) · · · 21.

The following statements hold for all n ≥ 3.

• Trajectory 12 · · · n has the highest probability and Trajectory n(n -1) · · · 1 has the lowest probability of all trajectories.

• Estimate 1 overestimates the probability of each substitution in Trajectory 12 · · · n, with exception for the first and the last step (compare Trajectory 12345).

• Estimate 1 underestimates the probability of each substitution in Trajectory n(n - 1) · · · 1, with exception for the first and the last step.

The proofs of these facts are elementary and of little theoretical interest since we are mostly interested in cases where n is relatively small, so we will omit the proofs. However, notice that the argument for the overestimate of trajectory A in Example 1 works in general. Indeed, according to Estimate 1 the available mutations constitute an $E_k$ set at any relevant step, whereas the fitness contributions of the genotypes we actually compare for Trajectory 12 · · · n are more equally distributed. The argument for the underestimates regarding Trajectory $n(n - 1) \cdots 1$ is similar.

The probabilities as well as the estimates for the extreme trajectories for n=4, 5, 6 can be found in Table 1. We found that estimate 2 gives results very similar to Estimate 1 for those extreme trajectories for n = 4, 5, 6. Table 2 shows the probabilities and estimates for every step of Trajectory 12345. Table 3 shows some trajectories where Estimates 1 and 2 give very different results.

*2.5 The accuracy of Estimate 1 and Estimate 2*

Throughout this study, we define the error of Estimate 1 as the ratio between the Estimate 1 probability and the exact probability of each mutational trajectory. The error for Estimate 2 is similarly defined. The error for Estimate 1 ranges between 0.52 and 3.24 for n=4, between 0.29 and 7.22 for n=5 and between 0.13 and 17.2 for n=6. The error for Estimate 2 ranges between 0.31 and 1.28 for n=4, between 0.14 and 1.73 for n=5, and between 0.12 and 1.9 for n=6.

As a comparison, Estimate 1 overestimates the probability by a factor 3 for n=4, a factor 7 for n=5 and a factor 17 for n=6 for particular trajectories. On the other hand, Estimate 2 underestimates the probability by a factor 3 for n=4, a factor 7 for n=5 and a factor 8 for n=6 for particular trajectories.

There are 24 trajectories for n = 4. When the 20 trajectories with the lowest probabilities according to Estimate 2 (these are the correct 20 trajectories) are considered, Estimate 2 underestimates the probabilities for 18 out of the 20 trajectories. The sum of the probabilities for these 20 trajectories is 35.9 % according to Estimate 2, whereas the correct value is 44.1 %. Estimate 2 overestimates the probabilities for each of the remaining 4 trajectories with the highest probabilities. The probabilities of each trajectory and the estimates are given in Table 4.

For n = 5, 99 out of 120 trajectories have a probability of at most 1% according to Estimate 2, which underestimates the probability for 96 out of the 99 trajectories. The sum of the probabilities of these 99 trajectories is 21.5 % according to Estimate 2, whereas the correct value is 32.5 %. On the other hand, the highest two probabilities are overestimated as 10.8% and 9.1 % by Estimate 2, when the correct values are respectively 8.1% and 6.1%. The tendencies of Estimate 1 for n = 4 and n = 5 are less clear. A density histogram of the log error distribution is shown in figure 1.

The log error distribution for both estimates for n=5 and n=6 are shown in figures 1 – 4. For Estimates 1 and 2 the mean $\mu$ and the SD are respectively shown in Tables 5 and 6. The fact that the $\mu$ is negative reflects that Estimate 2 underestimates the probabilities of most trajectories in the category with the lowest probabilities, where the threshold is such that the majority of trajectories falls into this category. (For the sake of completeness, we considered normality.

The distributions fail the Shapiro-Wilk test, as well as the normal quantile quantile plot test for normality, for Estimate 2 when n = 6 and for Estimate 1 for n=5 and n=6. The p-value of the Shapiro-Wilk test for Estimate 2 for n=5 is 0.28.)

*In summary, our results show that Estimate 1 and 2 perform poorly if fitness is additive. Under our assumptions, they overestimate the probability of the most likely trajectory, and underestimate the probability of the least likely trajectory. Perhaps most striking , Estimate 2 has a strong tendency to underestimate the probabilities of less common trajectories. As mentioned, for n = 5 the sum of the probabilities of the least likely 99 trajectories is 21.5 % according to Estimate 2, whereas the correct value is 32.5%. Clearly, Estimate 2 exaggerates the apparent predictability of evolution by underestimating the probabilities of less common trajectories.*

*2.6 Alternative approaches*

We will briefly sketch an alternative method for estimating reproductive fitness from fitness ranks, independent of the Orr-Gillespie approach. Readers not familiar with the terminology can omit this part without consequences for the further reading. Suppose that one wants to estimate probabilities for mutational trajectories for empirical fitness data. Moreover, assume that one has qualitative information about fitness only. More precisely, assume that the fitness ranks of the wild-type, an n-tuple mutant and all intermediates are known.

One needs a rough estimate of the fitness of the wild-type [such as 2 SD over the mean fitness], and of *the qualitative measure of additivity* [such as 0.75] (Crona et al. 2013a; Crona 2013). For some model of choice [such as rough Mt. Fuji fitness landscapes], find parameters so that the qualitative measure of additivity relative the wild-type agrees with the estimates [in the case suggested, the qualitative measure should be approximately 0.75 for a genotype with fitness 2 SD above the mean]. The choice of parameters constitutes a calibration of the landscape, which reflects additivity as well as the wild-type fitness. Next, one generates a large set of calibrated landscapes accordingly, and samples a subset of landscapes with the desired properties. Specifically, one wants a collection of $2^n$ genotypes which satisfy the following properties [in the case suggested]: The "wild-type" should have fitness 2 SD above the mean (one assigns a genotype with the right fitness as "the wild-type"). The remaining genotypes (*n* mutational neighbors, or "single mutants", as well as all genotypes combining the corresponding single mutations) should have fitness ranks which coincide with the data from the empirical example. Finally one computes the probability for each trajectory and landscape in the set, and determines the mean values. Those mean values constitute the estimate.

It remains to carefully test the approach we have sketched. Additivity is incorporated in a natural way for this estimate, in contrast to approaches based on the Orr-Gillespie theory.

## 3. Discussion

We have studied two estimates of probabilities for mutational trajectories based on extreme value theory. We have shown that these estimates give systematic errors for an additive fitness landscape in the sense that they overestimate the probability of the most likely trajectory, and underestimate the probability of the least likely trajectory. In addition, one of the estimates (Estimate 2) has a strong tendency to underestimate the probabilities of less common trajectories.

Any approaches, including simulations, which depend on similar arguments as the estimates will suffer from the same problems. Our result holds under the standard assumptions for this type of approaches, except that for the complication with additivity. More precisely, we assumed that the fitness spacing agrees with expected average behavior from extreme value theory. The fitness landscapes we consider in this article have a clear pattern that relatively few mutational trajectories correspond to a substantial part of the total probability. However, it is fair to say that Estimate 2 exaggerates the apparent predictability of evolution by underestimating the probabilities of less common trajectories. The complications for additive fitness landscapes are by no means surprising. However, approaches based on extreme value theory have been used in published work, and we are aware that journal reviewers have recommend this type of methods also for fitness landscapes with a substantial amount of additivity, which motivated this study. Our conclusion is that the estimates and similar arguments can be misleading in a situation with additivity. Estimate 2 is especially misleading in the sense that the estimate makes evolution appear more predictable than it is, because of the exaggerated differences of probabilities for trajectories. We wish to stress that the underlying complications will occur whenever one has more additivity than expected for an uncorrelated fitness landscape, including situations with epistasis and sign epistasis.

We have also sketched an alternative approach. Specifically, we propose an estimate for probabilities of mutational trajectories for fitness ranks which does not depend on the Orr-Gillespie approach.

From a broader perspective, anyone new to the field of fitness landscapes must ask why so much literature is devoted uncorrelated fitness. We appreciated that uncorrelated fitness is of interest as a theoretical extreme, but the literature seems somewhat excessive if motivated only as a theoretical exercise. At this point the empirical evidence is overwhelming that uncorrelated fitness is an unrealistic assumption. Nevertheless, there may have been some hope that results on uncorrelated fitness could work as approximations for realistic fitness landscapes. Our results suggest more caution. We have demonstrated that evolutionary predictability is sensitive for additivity. In our view, there are good reasons to believe that additivity strongly influences most aspects of adaptation, and has to be incorporated in the models.

REFERENCES

Aita, T. , Uchiyama, H., Inaoka,T., Nakajima M., Kokubo,T., Husimi, Y. (2000) Analysis of Local fitness landscape with a Model of the Rough Mt. Fuji- type Landscape: Application to Prolyl Endopeptidase and Thermolysin. Biopolyrrzers. 2000;54:64 -79.

Beerenwinkel, N., Eriksson, N. and Sturmfels, B. (2007). Conjunctive Bayesian networks. *Bernoulli 1*3:893–909.

Carnerio, M. and Hartl, D. L. (2010). Colloquium papers: Adaptive landscapes and protein evolution. Proc. Natl. Acad. Sci USA 107 suppl 1: 1747-1751.

Crona , K. (2013). Graphs, polytopes and fitness landscapes (book chapter), Recent Advances in the Theory and Application of Fitness Landscapes (A. Engelbrecht and H. Richter, eds.). Springer Series in Emergence, Complexity, and Computation.

Crona, K., Greene, D. and Barlow, M. (2013a). The peaks and geometry of fitness land- scapes. *J. Theor. Biol.* 317: 1–13.

Crona, K., Patterson, D, Stack, K, Greene, D., Goulart, C., Mahmudi, M, Jacobs, S., Kallman, M., Barlow, M. (2013b) *Antibiotic resistance landscapes: a quantification of theory-data incompatibility for fitness landscapes.* arXiv: 1303.3842

Desper, R., Jiang, F., Kallioniemi, O.P., Moch, H., Papadimitriou, C.H. and Schaffer, A.A., (1999). Inferring tree models for oncogenesis from comparative genome hybridization data. *Comput. Biol* 6 37-51.

Dobler, S., Dalla, S., Wagschal, V., Agrawal. A.A.. (2012). Community-wide convergent evolution in insect adaptation to toxic cardenolides by substitutions in the Na,K-ATPase.
Proc Natl Acad Sci U S A. 2012 Aug 7;109(32):13040-5. doi: 10.1073/pnas.1202111109.

Flyvbjerg, H. and Lautrup, B. (1992). Evolution in a rugged fitness landscape. Phys Rev A 46:6714-6723.

Goulart, C. P., Mentar, M., Crona, K., Jacobs, S. J., Kallmann, M., Hall, B. G., Greene D., Barlow M. (2013). Designing antibiotic cycling strategies by determining and understanding local adaptive landscapes. *PLoS ONE* 8(2): e56040. doi:10.1371/journal.pone.0056040.

Gillespie, J. H. (1983). A simple stochastic gene substitution model. *Theor. Pop. Biol.* 23 : 202–215.

Gillespie, J. H. (1984). The molecular clock may be an episodic clock. *Proc. Natl. Acad. Sci. USA 81* : 8009–8013.

Jorgensen, J. H. and Ferraro, M. J. (2009). Antimicrobial susceptibility testing: a review of general principles and contemporary practices. Clin Infect Dis:1749-55.

Kauffman, S. A. and Levin, S. (1987). Towards a general theory of adaptive walks on rugged landscapes. *J. Theor. Biol* 128:11–45.

Kingman, J. F. C. (1978). A simple model for the balance between selection and mutation. *J. Appl. Prob.* 15:1–12.

Kouyos, R. D., Leventhal, G. E., Hinkley, T., Haddad, M., Whitcomb, J. M., Petropoulos, C. J., & Bonhoeffer, S. (2012). Exploring the complexity of the HIV-1 fitness landscape. *PLoS genetics*, *8*(3), e1002551.

Kryazhimskiy S., Tkačik G., Plotkin J. B. (2009). The dynamics of adaptation on correlated fitness landscapes. Proc. Natl. Acad. Sci. USA 106, 18 638–18 643

Lobkovsky AE, Koonin EV. (2012). Replaying the tape of life: quantification of the predictability of evolution. Front Genet. 3:246. doi: 10.3389/fgene. 2012.00246.

Lobkovsky AE, Wolf YI, Koonin EV. (2011). Predictability of Evolutionary Trajectories in Fitness Landscapes. PLoS Comput Biol 7(12): e1002302. doi:10.1371/journal.pcbi.1002302

Macken, C. A. and Perelson, A.S, (1989) Protein evolution on rugged landscapes. *Proc. Nat. Acad. Sci.* USA 86:6192-6195.

Maynard Smith, J. (1970). Natural selection and the concept of protein space.

*Nature* 225:563–64.

Orr, H. A. (2002). The population genetics of adaptation: the adaptation of DNA sequences.*Evolution;* 56:1113–1124.

Orr, H. A. (2006). The population genetics of adaptation on correlated fitness landscapes: the block model. *Evolution;* 60:1317–1330.

Ostman, B.and Adami (2013) Recent Advances in the Theory and Application of Fitness Landscapes" (A. Engelbrecht and H. Richter, eds.). Springer Series in Emergence, Complexity, and Computation.

Park, S. C. and Krug J. (2008). Evolution in random fitness landscapes: The infinite sites model. J Stat Mech P04014.

Rokyta, D. R., Beisel, C. J. and Joyce, P. (2006) Properties of adaptive walks on uncorre- lated landscapes under strong selection and weak mutation. *J Theor. Biol.* 2006;243:114.

Schenk, M. F,,  de Visser, J. A., Predicting the evolution of antibiotic  resistance BMC Biol. 2013 Feb 22;11:14. doi: 10.1186/1741-7007-11-14.

Szendro, IG, Franke, J, de Visser JA, Krug, J. (2013a).Predictability of evolution dependsnonmonotonically on population size. Proc Natl Acad Sci U S A. Jan 8 110(2):571-6

Szendro, I. G., Schenk, M. F., Franke, J. Krug, J. and de Visser J. A. G. M. (2013b). Quantitative analyses of empirical fitness landscapes. *J. Stat. Mech.* P01005.

Weinreich, D. M., Delaney N. F., Depristo, M. A., and Hartl, D. L. (2006). Darwinian evolution can follow only very few mutational paths to fitter proteins. *Science 312:*111–114.

Wright, S. (1931). Evolution in Mendelian populations. *Genetics,* 16 97–159.

**Table 1. Probabilities of trajectories**

| Extreme Trajectories | | | | | | |
|---|---|---|---|---|---|---|
| | N=4 | | N=5 | | N=6 | |
| Trajectory | 1234 | 4321 | 12345 | 54321 | 123456 | 654321 |
| Probability | 0.206 | 0.00333 | 0.0817 | 0.000243 | 0.0287 | $1.43 \cdot 10^{-5}$ |
| Estimate 1 | 0.239 | 0.00174 | 0.109 | $6.94 \cdot 10^{-5}$ | 0.0445 | $1.93 \cdot 10^{-6}$ |
| Estimate 2 | 0.264 | 0.00173 | 0.108 | $7.71 \cdot 10^{-5}$ | 0.0420 | $2.43 \cdot 10^{-6}$ |
| Probabilities for each step of Trajectory 12345† | | | | | | |
| | Step 1 | Step 2 | Step 3 | Step 4 | Step 5 | |
| Probability | 0.457 | 0.472 | 0.547 | 0.692 | 1 | |
| Estimate 1 | 0.457 | 0.521 | 0.611 | 0.750 | 1 | |
| Estimate 2 | 0.548 | 0.494 | 0.557 | 0.714 | 1 | |
| Probabilities where the estimates differ considerably | | | | | | |
| Trajectory | Trajectory 2341 for n=4 | | Trajectory 23451 for n = 5 | | Trajectory 234561 for n = 6 | |

| Probability | 0.00580 | 0.000668 | $6.77 \cdot 10^{-5}$ |
| Estimate 1 | 0.0188 | 0.00483 | 0.00117 |
| Estimate 2 | 0.00181 | 0.000114 | $1.03 \cdot 10^{-5}$ |

†Notice that Estimate 1 gives an overestimate of every step of Trajectory 12345, except for Step 1 and the very last step.

**Table 2. Log error mean and standard deviation**

| n | 4 | 5 | 6 |
|---|---|---|---|
| Estimate 1 | | | |
| μ | 0.090 | 0.157 | 0.231 |
| s | 0.570 | 0.760 | 0.942 |
| Estimate 2 | | | |
| μ | -0.493 | -0.921 | -1.090 |
| s | 0.570 | 0.813 | 0.904 |

**Table S1. Probabilities for all trajectories for n=4**

| Trajectory | Probability | Estimate 1 | Estimate 2 |
|---|---|---|---|
| 1234 | 0.2060700 | 0.2387200 | 0.2643200 |
| 1243 | 0.0883150 | 0.0795720 | 0.1057300 |
| 1324 | 0.1287900 | 0.1085100 | 0.1304700 |
| 1342 | 0.0297210 | 0.0361690 | 0.0203330 |
| 1423 | 0.0441580 | 0.0434030 | 0.0502190 |
| 1432 | 0.0237770 | 0.0144680 | 0.0193830 |
| 2134 | 0.1354200 | 0.1241300 | 0.1401800 |
| 2143 | 0.0580360 | 0.0413770 | 0.0560710 |
| 2314 | 0.0483630 | 0.0564240 | 0.0297670 |
| 2341 | 0.0058036 | 0.0188080 | 0.0018084 |
| 2413 | 0.0181360 | 0.0225690 | 0.0131010 |
| 2431 | 0.0050781 | 0.0075231 | 0.0019354 |
| 3124 | 0.0722500 | 0.0668400 | 0.0658300 |
| 3142 | 0.0166730 | 0.0222800 | 0.0102590 |
| 3214 | 0.0412860 | 0.0303820 | 0.0280880 |
| 3241 | 0.0049543 | 0.0101270 | 0.0017064 |
| 3412 | 0.0070202 | 0.0121530 | 0.0047368 |
| 3421 | 0.0036505 | 0.0040509 | 0.0017898 |
| 4123 | 0.0225690 | 0.0286460 | 0.0246640 |
| 4132 | 0.0121530 | 0.0095486 | 0.0095194 |
| 4213 | 0.0141060 | 0.0130210 | 0.0119860 |
| 4231 | 0.0039497 | 0.0043403 | 0.0017708 |
| 4312 | 0.0063962 | 0.0052083 | 0.0045917 |
| 4321 | 0.0033260 | 0.0017361 | 0.0017349 |

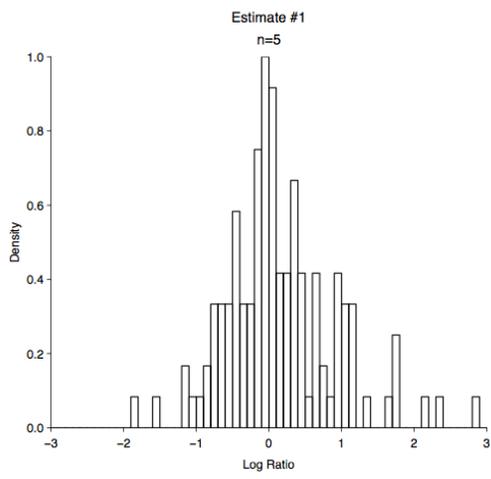 A

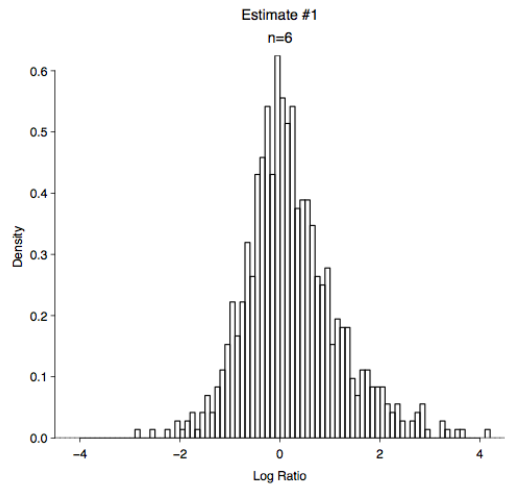 B

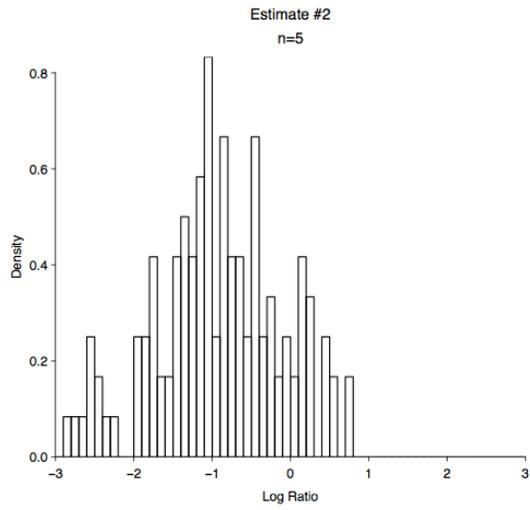 C

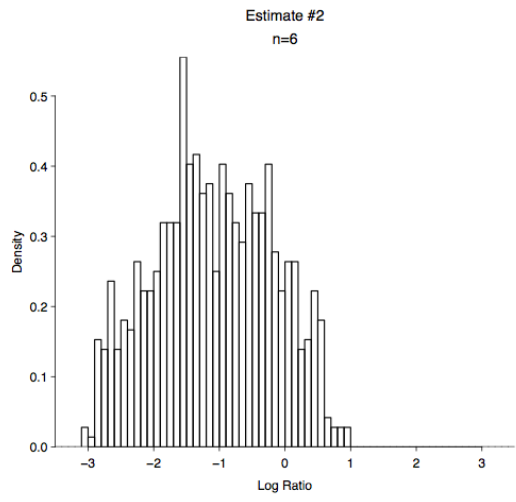 D

33